\documentclass[english,aps, superscriptaddress, floatfix, 12pt,
  nofootinbib]{revtex4} 

\usepackage{graphicx}
\usepackage{amssymb}
\usepackage{amsmath}

\makeatletter

\providecommand{\tabularnewline}{\\}

\newcommand{\be}{\begin{eqnarray}}
\newcommand{\ee}{\end{eqnarray}}

\newcommand{\bea}{\begin{eqnarray}}
\newcommand{\eea}{\end{eqnarray}}

\newcommand{\E}{\mathrm{e}}

\def\slash#1{\setbox0=\hbox{$#1$}               
   \dimen0=\wd0                                 
   \setbox1=\hbox{/} \dimen1=\wd1               
   \ifdim\dimen0>\dimen1                        
      \rlap{\hbox to \dimen0{\hfil/\hfil}}      
      #1                                        
   \else                            
      \rlap{\hbox to \dimen1{\hfil$#1$\hfil}}   
      /                                         
   \fi}                                         %
   
\usepackage{babel}
\makeatother

\begin{document}

\title{Effects of the Running of the QCD Coupling\\ 
on the Energy Loss in the Quark-Gluon Plasma}

\author{Jens Braun}
\affiliation{Institute for Theoretical Physics, University of
  Heidelberg, Philosophenweg 19, 69120 Heidelberg, Germany}
\author{Hans-J\"urgen Pirner}
\affiliation{Institute for Theoretical Physics, University of
  Heidelberg, Philosophenweg 19, 69120 Heidelberg, Germany}
\affiliation{Max-Planck-Institut f\"ur Kernphysik, Saupfercheckweg 1,
  69117 Heidelberg, Germany}

\begin{abstract}
Finite temperature modifies the running  of the 
QCD coupling $\alpha_s(k,T)$ with re\-solution $k$.
After calculating  the thermal quark and gluon masses  selfconsistently, we
determine the quark-quark and 
quark-gluon cross sections in the plasma based on
the running coupling. We find that the running coupling enhances these cross sections  
by factors of two to four depending on the temperature. 
We also compute the energy loss $\frac {d E}{dx}$
of a high-energy quark in the plasma as a function of temperature. Our study suggests that, 
beside $t$-channel processes, inverse Compton scattering is a relevant process for a quantitative 
understanding of the energy loss of an incident quark in a hot plasma.
\end{abstract}

\maketitle

\bigskip
\section{Introduction}
The consequences of a strongly interacting 
phase of QCD at high temperatures are currently very actively investigated. 
The existence of such a phase was suggested in Ref.~\cite{Shuryak:2003xe} 
after a careful analysis of QCD lattice data.
Moreover, it has been known for quite a while that perturbative approaches to QCD 
cannot give final answers to many questions of heavy-ion 
collision experiments, such as the short thermalization times of the quark-gluon 
plasma.

High-temperature effective field theory has often 
used a running QCD coupling depending on the temperature only 
which may be justified at very high 
temperatures~$T\gg \Lambda _{\text{QCD}}$, see e.g.~\cite{Blaizot:2003tw}.
In current heavy-ion collision experiments, however, 
the temperature $T$ may not be the only important 
scale, it is one among others, such as the momenta $p$ of the partons. 
Therefore the running of the coupling 
in the momentum regime, where $p\sim T$, has to be taken into account for the 
computation of cross sections and energy-loss processes. In this paper 
we present a first approach in this direction.
In Sec.~\ref{sec:RGcoup}, we review recently obtained results from 
a calculation of the 
strong coupling from first principles within the framework of the 
functional RG~\cite{Gies:2002af,Braun:2005uj,Braun:2006jd}. Moreover,
we deduce a parametrization for strong coupling at zero and finite temperature 
from these results. We use the results from Refs.~\cite{Braun:2005uj,Braun:2006jd} 
for a computation of the parton cross section and the energy loss of partons in 
the quark-gluon plasma in Sec.~\ref{sec:cross_section} and~\ref{sec:jet}, respectively. 
In particular, we study the energy loss of an incident quark in the quark-gluon plasma 
incorporating $t$-channel processes and inverse Compton scattering. 
Taking the running coupling of QCD into account, we find 
that the energy loss of an incident quark in a hot plasma is increased. Moreover,
our results suggest that inverse Compton scattering 
gives a third of the energy loss of an incident quark. 
Our conclusions and future plans are found in Sec.~\ref{sec:conc}.
\section{Running of the strong coupling}\label{sec:RGcoup}
Recently the QCD coupling $\alpha_s(k,T)$ at
finite temperature has been calculated by nonperturbative 
functional Renormalization Group methods \cite{Braun:2005uj} and shows 
a considerable variation with the resolution $k$.
The calculation employs the background-field formalism \cite{Abbott:1980hw} within 
the RG framework \cite{Gies:2002af,Reuter:1993kw}.
The effective action used for the determination of the running coupling 
includes an infinite power series of the gauge-invariant operator 
$F^a_{\mu\nu}F^a_{\mu\nu}$:
\be
\Gamma _k= \int _x \left\{\sum_{i} Z_k ^{(i)} \left(F_{\mu \nu}^a
F_{\mu \nu}^a\right)^i + \bar{\psi}\slash{D}[A]\psi\right\} +
\text{gauge-fixing term} + \text{ghost term}.\label{eq:trunc}
\ee
The truncation of the full effective action includes arbitrarily
high gluonic correlators projected onto their small-momentum limit and
onto the particular color and Lorentz structure arising from powers of
$F^a_{\mu\nu}F^a_{\mu\nu}$. It represents a gradient expansion in the field strength 
and neglects higher-derivative terms and more complicated color and
Lorentz structures. So far possible differences between 
magnetic and electric terms have not been considered. In the background-field method, the 
$\beta$-function of the running coupling $g$ is related to wave function renormalization of the background field \cite{Abbott:1980hw} via
\begin{eqnarray}
 k\frac{\partial}{\partial k}g^2 = \eta\,g^{2}\qquad\text{with}\qquad
 \eta =-k\frac{\partial}{\partial k} \ln Z_k ^{(1)}\,,
\label{betadef}
\end{eqnarray}
as a consequence of the non-renormalization of the product 
of the background field and the coupling. The coefficient of the 
first term  $Z_k ^{(1)}$ in the effective action \eqref{eq:trunc} 
evolves with the renormalization scale $k$, successively driven by all 
other operators in the action. 
In Refs. \cite{Gies:2002af,Braun:2005uj} the 
authors keep track of all contributions from the flows of the $Z_k ^{(i)}$ 
to the flow of the running coupling. 
They use as initial condition for the coupling the measured value 
at the $\tau$-mass scale \cite{Bethke:2004uy}, $\alpha_{\mathrm{s}}=0.322$.
The resulting $\beta$-function at zero temperature agrees well with the
perturbative two-loop running in the $\overline{\text{MS}}$-scheme 
at high resolution $k$.
It approaches a strong-coupling fixed point in the infrared (IR) \cite{Gies:2002af} in 
accordance with results from the Schwinger-Dyson equations and RG flow equations
in Landau gauge \cite{RG_Landau_gauge,DSE_Landau_gauge}.
At zero temperature, the coupling $\alpha_s$ can be parametrized by
\be
\alpha _{s}(k,T=0)=
\frac{\alpha _{s} ^{*}}{\ln \left(e + \left(\frac{k}{\Lambda _{s}}\right)^a
 + \left(\frac{k}{\Lambda _{s}}\right)^b\right)}\,,
\label{Eq:alpha_T0}
\ee
where $\alpha _{s} ^{*}=5.7$ denotes the value of the IR fixed point and $e=2.718$. 
The parameters $a,b$ and $\Lambda _s$ are given by
\be
a=9.07,\quad b=5.90\quad\text{and}\quad\Lambda _s=0.263\,\text{GeV}\,.
\label{Eq:para_T0}
\ee
Similar parametrizations were also found in studies in terms of  
Schwinger-Dyson equation~\cite{DSE_Landau_gauge}.

\begin{figure}
\includegraphics[clip,scale=0.9]{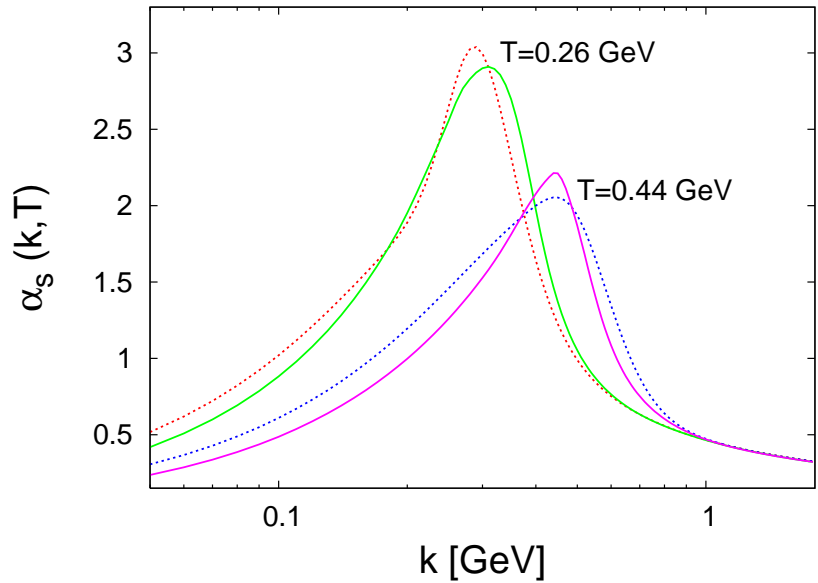}
\caption{\label{fig:alpha_plot} Comparison of the fit of the 
running coupling (dotted lines) obtained from Eqns. \eqref{Eq:alpha_T} and \eqref{eq:fitpara} 
with the full numerical result (straight line) for pure Yang-Mills theory 
from Refs.~\cite{Braun:2005uj,Braun:2006jd} for $T=0.26\;\text{GeV}$ and $T=0.44\;\text{GeV}$.} 
\end{figure}

At finite temperature $T$, the UV behavior remains unaffected
for scales $k\gg T$ and agrees well with the perturbative running
coupling at zero temperature for high-momentum scales.  
In the infrared, however, the running is strongly modified compared to results 
for vanishing temperature:
$\alpha_s(k,T)$ reaches a maximum at $k\sim T$ and then 
decreases towards lower scales, see Fig.~\ref{fig:alpha_plot}. 
This can be easily understood in terms of a simple picture: 
below the scale $k\sim T$, gluonic fluctuations with a Compton wave length 
larger than the extent of the Euclidean time direction 
become effectively three-dimensional. Therefore the limiting behavior of 
the coupling flow in this regime is governed by the spatial 3$d$ Yang-Mills 
theory. Nontrivial, however, is the fact that the coupling decreases for $k<T$. 
This is due to the existence of a non-Gaussian IR fixed point in the 
three-dimensional theory \cite{Braun:2005uj,Braun:2006jd}.
In the low momentum regime, the solution of the RG-equations gives 
a linear behavior with a slope determined by the infrared fixed point of 
the spatial 3$d$ Yang-Mills theory \cite{Braun:2006jd}:   
\be
\alpha _s (k\ll T)\approx\alpha ^{*} _{3d}\,\frac{k}{T}
+{\mathcal O}\left(\left(\frac{k}{T}\right)^2\right)\,.
\label{Eq:limit_3d}
\ee
It is well known that the low-momentum limit
of thermal loops enhances the powers of the coupling as $\alpha_s T/k$. 
From Eq.~\eqref{Eq:limit_3d}, we can read off that also the 
enhanced effective coupling  $\alpha_s T/k$ remains constant for $k\rightarrow 0$, 
similarly to the coupling $\alpha_s$ at zero temperature.
We note that the IR fixed point in the 3$d$ Yang-Mills theory is 
also in accordance with recent results in the Landau gauge \cite{Maas:2004se}.

A good parametrization of the  running coupling at finite temperature 
which respects the perturbative UV behavior 
and the 3$d$ IR fixed point has the following form:
\be
\alpha _{s}(k,T)=
\frac{u_{1}\,\frac{k}{T}}{1+\exp \left(u_{2} \frac{k}{T} - u_{3}\right)}+
\frac{v_{1}}{\left(1+\exp \left(v_{2} \frac{T}{k} - v_{3}\right)\right)\,
\ln \left(e + \left(\frac{k}{\Lambda _{s}}\right)^a
 + \left(\frac{k}{\Lambda _{s}}\right)^b\right)}\,.
\label{Eq:alpha_T}
\ee
If we choose the parameters $a,\,b$ and $\Lambda _s$ to be the same as in
Eq.~\eqref{Eq:para_T0}, then the above parametrization agrees with this
equation at $T=0$ for
\be
v_{1}=\alpha _{s} ^{*}\,(1+\exp\left(-v_{3}\right))\,.
\ee
In order to take the limiting behavior of the coupling for $k\ll T$
into account, we choose
\be
u_{1}=\alpha ^{*} _{3d}\,\left(1+\exp\left(-u_{3}\right)\right)\,.
\ee
Here $\alpha ^{*} _{3d}$ and $\alpha _{s}^{*}$ denote the values of the IR fixed point 
of $SU(3)$ Yang-Mills theory in $d=3$ and $d=4$ dimensions, respectively.
The remaining four parameters fit the numerical results for pure Yang-Mills theory 
obtained from the RG equations in Ref. \cite{Braun:2005uj}:
\be
u_{2}=5.47,\quad u_{3}=6.01,\quad v_{2}=10.13\quad\text{and}\quad v_{3}=9.27\,.
\label{eq:fitpara}
\ee
As shown in Fig. \ref{fig:alpha_plot}, the fit is in reasonable agreement 
with the results of Ref. \cite{Braun:2005uj}
for temperatures $T\in~[0.2\,\text{GeV},0.6\,\text{GeV}]$.

Phenomenologically, the strong variation of the running coupling as a
function of the resolution $k$ is important. 
Especially the behavior of the coupling near its maximum
value influences physical observables. With increasing temperature the
position of the maximum shifts to higher momenta and the value at the
maximum decreases. On average, the system becomes less
strongly coupled for higher temperature, in agreement with naive
expectations from a temperature-dependent effective coupling. This
behavior influences the scattering cross section of a quark with
the particles in the quark-gluon plasma. As we will discuss below, 
this elastic cross section contributes significantly to the energy loss of 
an incident quark in the quark-gluon plasma. 

\section{Effective cross section of an incident quark\label{sec:cross_section}}

\begin{figure}
\includegraphics[clip,scale=0.9]{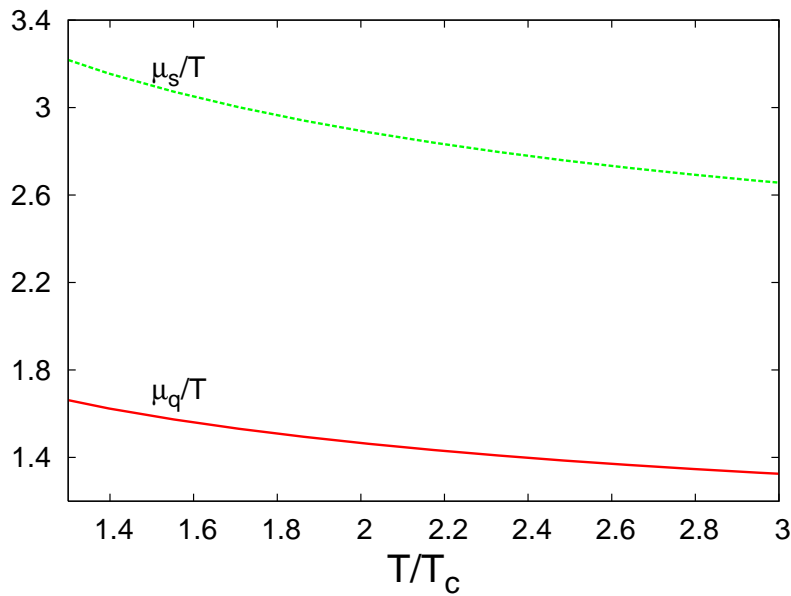}
\caption{\label{fig:mu_plot} The figure shows the dimensionless 
ratios of the gluon screening mass $\mu _s$ and 
the quark mass $\mu_q$ over temperature $T$ as a function of the dimensionless 
quantity $T/T_c$ for $N_f =3$. 
The chiral phase transition temperature $T_c$ is given in Eq.~\eqref{eq:Tc}.}
\end{figure}

Before we turn to the calculation of the cross sections, we discuss the
Debye mass $\mu _{s}$ which enters as screening mass in this
calculation. 
As argued in Ref.~\cite{Peshier:2006ah}, 
the relevant coupling for the calculation of the Debye mass $\mu _s$ 
must be fixed self-consistently at the scale $k=\mu_s$, since in the 
gluon polarization operator external
gluons have vanishing momentum. Therefore we use 
the following self-consistent equation for the determination of the Debye mass $\mu _s$
to leading order~\cite{Peshier:2006ah}:
\begin{equation}
\mu_{s} ^2 (T)= \frac{4\pi}{3}N_{c}\left(1+\frac{1}{6}N_f\right)
\alpha_{s}  (\mu_{s}(T),T)\, T^2\,.
\label{Eq:self_const}
\end{equation}
Results for the Debye mass $\mu _s$
are shown in Fig.~\ref{fig:mu_plot} as a function of temperature for
three flavors.  
Here and in the following we identify the scale $k^2$ 
set by the RG with the momentum scale $p^2$. This nontrivial assumption  
is justified, because the regulator function entering 
in the calculation of the running coupling specifies the Wilsonian momentum-shell 
integration in such a way that the RG flow of the coupling is dominated by fluctuations 
with momenta $p^2 \simeq k^2$.

Since we are interested in  temperatures above the critical temperature, 
the relevant values of the running coupling $\alpha_{s} ^2(\mu_{s}(T),T)$ 
in Eq. \eqref{Eq:self_const} lie near the asymptotic two-loop 
scaling region, i.e. above the scale set by temperature $T$. 
The mass of the renormalized quark propagator in hot QCD was calculated 
in Ref. \cite{Pisarski:1987wc}. Along the lines of our calculation of the 
Debye mass, we use this result for a calculation of an "effective quark mass" $\mu _q$ 
by solving the corresponding self-consistent equation:
\begin{equation}
\mu_{q} ^2 (T)= \frac{\pi}{2}\frac{N_{c}^2 -1}{2N_c}\alpha_{s}(\mu_{q}(T),T)\,T^2\,.
\label{Eq:quark_self_const}
\end{equation}
The results for $\mu _s (T)$ as well as $\mu _q (T)$ are shown in 
Fig.~\ref{fig:mu_plot} for $N_f =3$ massless quark flavors.
The solutions of the self-consistency equations~\eqref{Eq:self_const} 
and~\eqref{Eq:quark_self_const} 
exhibit a behavior in form of an alternating series in powers of $\frac{T}{T_c}$:
\be
\frac{\mu _{j}(T)}{T} \approx m_{0}^{(j)} - m_{1}^{(j)} \left(\frac{T_c}{T}\right)+
m_{2}^{(j)} \left(\frac{T_c}{T}\right)^2 -
m_{3}^{(j)} \left(\frac{T_c}{T}\right)^3 +  ...\,,\quad\text{with}
\quad j\in \{q,g\}.
\label{Eq:fit_mu}
\ee
For our analysis, we use the following chiral phase transition temperatures 
\be 
T_c(N_f=2)\approx 0.186\,\text{GeV}\quad\text{and}\quad
T_c(N_f=3)\approx 0.161\,\text{GeV}\,.\label{eq:Tc}
\ee
These values, which were obtained from an RG calculation \cite{Braun:2006jd}, 
are in good agreement with lattice QCD studies \cite{Karsch:2000kv}. 
The coefficients $m_{i}^{(j)}$ in Eq.~\eqref{Eq:fit_mu} depend on the number of flavors $N_f$. 
The values of these coefficients for a truncated series, including powers up to 
$(T/T_c)^3$, obtained from a fit to the full numerical data are 
given in Tab.~\ref{Tab:coeff}. 
\begin{table}
\begin{ruledtabular}
\begin{tabular}{c||cccc}
$i$&
0&
1&
2&
3\tabularnewline
\hline
\hline 
$m_{i}^{(g)}(N_{f}=2)$&
4.31&
-1.45&
0.40&
-0.04\tabularnewline
\hline 
$m_{i}^{(g)}(N_{f}=3)$&
4.29&
-1.15&
0.28&
-0.02\tabularnewline
\hline 
$m_{i}^{(q)}(N_{f}=2)$&
2.50&
-0.96&
0.27&
-0.03\tabularnewline
\hline 
$m_{i}^{(q)}(N_{f}=3)$&
2.32&
-0.70&
0.17&
-0.02\tabularnewline
\end{tabular}
\end{ruledtabular}
\caption{The fitted values for the coefficients $m_{i}^{(j)}$ of the
power series~\eqref{Eq:fit_mu} for $N_f =2$ and $N_f =3$.\label{Tab:coeff}}
\end{table}

The Debye mass as well as the ``effective quark mass" enter into the calculation of the
effective cross section which a fast quark experiences when it traverses the quark-gluon
plasma. These masses serve as an infrared cut-off for the  cross section. 
In the following, we neglect the spins of the partons and
consider the cross section averaged over initial and summed over final 
colors of both projectile and target partons. In a first approach, 
this is a justified assumption for the processes under consideration \cite{Gyulassy:2003mc}.
Using the temperature-dependent coupling $\alpha_s(p,T)$, the Debye mass $\mu_s (T)$ 
as well as the ``effective quark mass" $\mu _q (T)$, 
the total cross section of an incident quark obtains
\be
\sigma _{i}(T)= C_{i}\int _0 ^{s}
d|t|\,\frac{d\sigma ^{t-\text{ch.}}}{d|t|} + \delta _{i(qg)}\int _{\mu _q ^2 (T)} ^{s}
d|u|\,\frac{d\sigma ^{\text{IC}}}{d|u|}\qquad\left(i\in\{qq,qg\}\right)\,,\label{Eq:sigma}
\ee
where the factor $C_i$ denotes the respective color factor which 
is $C_{qq}=4/9$ for $qq$-scattering and $C_{qg}=1$ for $qg$-scattering. 
The differential cross sections for $t$-channel and $u$-$s$-channel processes 
are given by \cite{Gyulassy:2003mc,Ellis:1991qj}
\be
\frac{d\sigma ^{t-\text{ch.}}}{d|t|}
=\frac{2\pi\alpha_s ^2\left(\sqrt{|t|},T\right)}{\left(|t|+\mu_s ^2(T)\right)^2}
\quad\text{and}\quad\frac{d\sigma ^{\text{IC}}}{d|u|}=\frac{4\pi}{9 s^2}
 \,\left(\alpha _s ^2 \left(\sqrt{s},T\right)\frac{|u|}{s} +\alpha _s ^2 \left(\sqrt{|u|},T\right)\frac{s}{|u|}\right).
\label{Eq:dsdt}
\ee
In the  calculation of the quark-gluon cross section we take into account $t$-channel exchange processes as well as u-s-channel exchange processes, 
where the latter ones are the only processes possible in QED. 
They are denoted as inverse Compton-scattering (IC scattering) processes, 
where a fast fermion hits a thermal boson. 
In high-energy astrophysics the energy loss of electrons in our galaxy 
is dominated by scattering on photons of the cosmic-microwave background \cite{Bere}.
Moreover, it is also the main process for the production of high-energy 
$\gamma$-rays \cite{Aharonian}. 
\begin{figure}
\includegraphics[clip,scale=0.9]{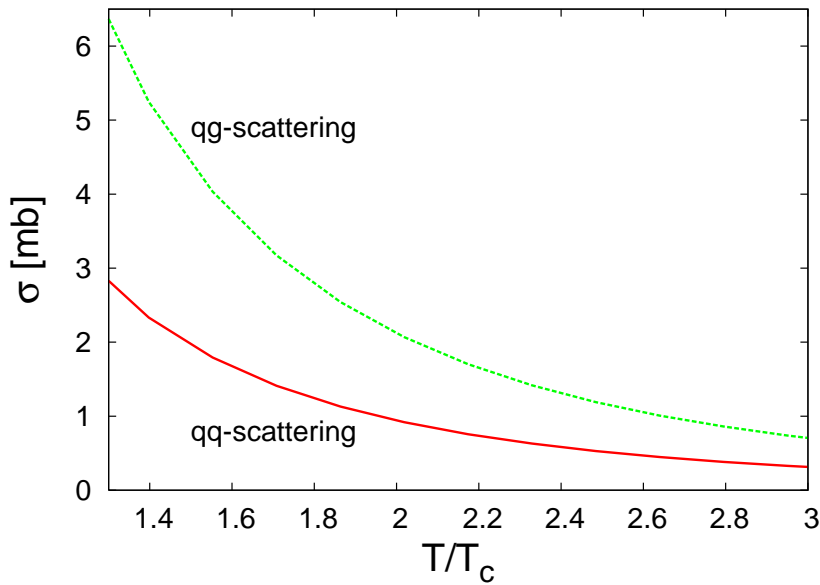}
\caption{\label{fig:cross_plot}The cross sections  
for qq and qg-scattering in the quark-gluon plasma are 
shown as a function of the dimensionless 
quantity $T/T_c$ for $N_f=3$ massless quark flavors in the limit 
$s\rightarrow\infty$. The chiral 
phase-transition temperature is given by $T_c\approx 161\;\text{MeV}$.} 
\end{figure}

At finite temperatures, the effective differential cross section is 
enhanced for $p\approx T$ due to the increasing of the running coupling with 
momentum transfer $p$. This in turn 
leads to a larger integrated  cross section $\sigma _{i}$ itself.
The numerical results for the integrated cross sections  are given 
in Tab.~\ref{Tab:cross section}. To be specific, we show the cross section 
for $qq$-scattering and $qg$-scattering in Fig.~\ref{fig:cross_plot}.
If one uses the running QCD coupling of Ref. \cite{Braun:2005uj} 
for three massless quark flavors, one
gets a $qq$ cross-section of $\sigma _{qq}(T=1.3\,T_c)\approx 2.79\,\text{mb}$ 
at $T=1.3\,T_c$ and of $\sigma _{qq}(T=3.3\,T_c)\approx 0.24\,\text{mb}$ 
at $T=3.3\,T_c$.

For comparison, we have added in Tab.~\ref{Tab:cross section} the
calculation of the $qq$-cross section if the strong coupling would be
constant $\alpha_s=0.5$. The result from our 
improved calculation is enhanced by a factor of four maximally
compared with the calculation using constant $\alpha_s$.
\begin{table}[t!]
\begin{ruledtabular}
\begin{tabular}{c||ccccc}
$\frac{T}{T_c}$&
1.3&
1.8&
2.3&
2.8&
3.3\tabularnewline
\hline
\hline 
$\sigma_{qg}(T,N_{f}=3)\,\,\text{[mb]}$&
6.29&
2.78&
1.46&
0.86&
0.54\tabularnewline
\hline 
$\sigma_{qq}(T,N_{f}=3)\,\,\text{[mb]}$&
2.79&
1.23&
0.65&
0.38&
0.24\tabularnewline
\hline 
$\sigma_{qq}(T,N_{f}=3,\alpha_s = 0.5)\,\,\text{[mb]}$&
0.66&
0.34&
0.21&
0.14&
0.10\tabularnewline
\end{tabular}
\end{ruledtabular}
\caption{\label{Tab:cross section}Total cross section calculated 
from Eq. \eqref{Eq:sigma} in the limit $s\rightarrow \infty$ 
for different values of $T/T_c$ for $qq$-scattering and $qg$-scattering 
of the incident quark. For comparison we also show the results for $\sigma_{qq}(T)$ 
for $\alpha_s (k,T)=\alpha _s =0.5$. 
The chiral phase-transition temperature is given by $T_c\approx 161\;\text{MeV}$.}
\end{table}
So the mean free path with the running coupling becomes much shorter
and even elastic collisions may enter into the energy loss
scenario.

\section{Jet's collisional energy loss}\label{sec:jet}

In this section, we discuss the collisional energy loss of an incident quark
in a hot plasma which we study with the running coupling depending on 
temperature and resolution. Ground-breaking work on energy loss 
of energetic partons in a hot plasma was done by Bjorken~\cite{Bjorken:1982tu}. 
Recently this calculation has been redone with a running coupling in one-loop approximation 
independent of temperature~\cite{Peshier:2006hi}. 
We are interested in the energy loss due to elastic scattering of a high-energy 
quark from a parton of momentum $p$ in the plasma for temperatures $T>\Lambda _{\text{QCD}}$. 
Let us start with the calculation including the cross section for $t$-channel exchange 
processes ($t\ll s$). 

\begin{figure}
\includegraphics[clip,scale=0.9]{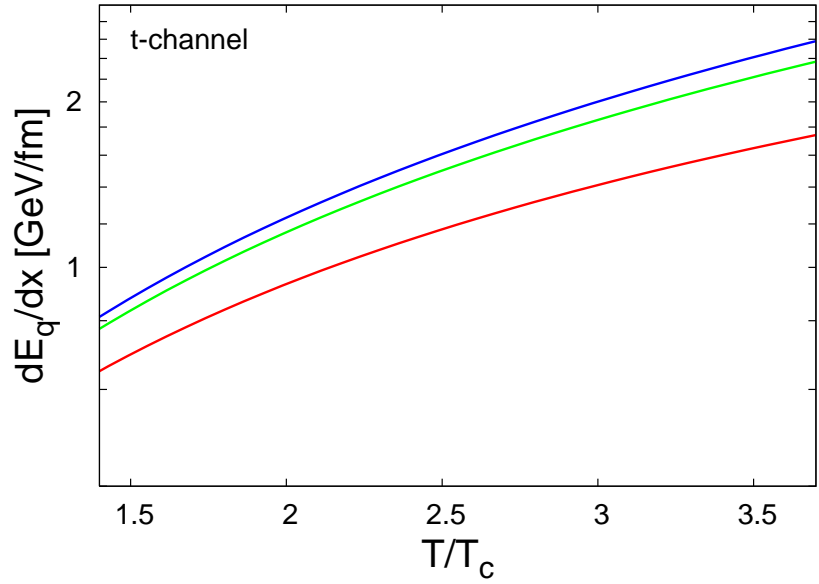}
\caption{\label{fig:dEdx_plot} Collisional energy loss of a massless
quark in QCD with $N_f =3$ massless quark flavors due to $t$-channel 
gluon exchange as a function of $T/T_c$. We show the results for energies 
$E_q=20,\,100,\,200\,\text{GeV}$ of the incident quark (from bottom to top).
The chiral phase transition temperature is given by $T_c\approx 161\;\text{MeV}$.}
\end{figure}

The cross section for such processes is given in Eq. \eqref{Eq:dsdt}.
The energy loss $dE$ of the high-energy quark per length $dx$ in 
the hot plasma due to $t$-channel processes can be calculated as~\cite{Bjorken:1982tu}
\be
\frac{dE _q ^{t-\text{ch.}} (T)}{dx}=\frac{2}{3}\int d^3 p\, \rho_{\text{eff.}}(p,T)
\,\Phi\int _{t_1}^{t_2} dt \frac{d\sigma ^{\text{t-ch.}}}{dt}\nu (p,t)\,.
\label{eq:dEdx1}
\ee
Here $\Phi$ denotes the flux factor and $\nu =E-E'$ is the energy difference 
of the incident and emergent parton depending on  the thermal momentum $p$ and 
the momentum transfer~$t$. 
Considering the Casimir factors for the various processes, we have defined
an effective plasma density $\rho _{\text{eff.}}(k,T)$:
\be
\rho _{\text{eff.}}(p,T)=\frac{2}{3}\,\rho_{q}(p,T) + \frac{3}{2}\,\rho_{g}(p,T)\,,
\ee
where $\rho _{q}(p,T)$ and $\rho _{g}(p,T)$ are the quark and gluon momentum distributions,
respectively:
\be
\rho_{q}(p,T)=\frac{12 N_f}{(2\pi)^3}\frac{1}{\E ^{|p|/T} +1}\quad\text{and}\quad
\rho_{g}(p,T)=\frac{16}{(2\pi)^3}\frac{1}{\E ^{|p|/T} -1}\,.
\label{eq:rho}
\ee
In our calculation, we take into account that the coupling $\alpha_s(t,T)$
depends on both momentum transfer $t$ and temperature $T$. 
Finally, the energy loss will be a function of the 
temperature and incident energy. 
Note that the running of $\alpha _s$ was not taken into account by Bjorken 
and only partially taken into account by a temperature-independent one-loop 
approximation in Ref.~\cite{Peshier:2006hi}.
Following Ref.~\cite{Bjorken:1982tu}, we can write $\nu$ for $E,\,E'\gg |p|$ as
\be
\nu=\frac{|t|}{2p\Phi}\,.\label{eq:nu}
\ee
Using Eqns. \eqref{Eq:dsdt},\,\eqref{eq:rho} and \eqref{eq:nu}, the energy loss 
of an incident quark propagating through a hot plasma is determined by
\be
\frac{dE_q ^{t-\text{ch.}}(T)}{dx}=\frac{4}{3}\,\pi \left(1+\frac{N_f}{6}\right) T^2
\int _{|t_2|} ^{|t_1|} d|t|\, 
\frac{\alpha _s ^2 (\sqrt{|t|},T)}{\left(|t|+\mu_s ^2(T)\right)^2}\,|t|\,.
\ee
In our calculations, we use $|t_1|=s$ and $|t_2|=0$. Note that the energy of the incident quark~$E_q$ and the average value of $s$ are related by $s=2E_{q}T$.
However, even the limit $|t_1|\rightarrow\infty$ would be well-defined since 
the coupling decreases logarithmically. In Fig. \ref{fig:dEdx_plot} we show the 
energy loss of a massless quark as a function 
of $T/T_c$ for different energies $E_q$ of the incident quark 
due to $t$-channel exchange processes. 
\begin{figure}
\includegraphics[clip,scale=0.9]{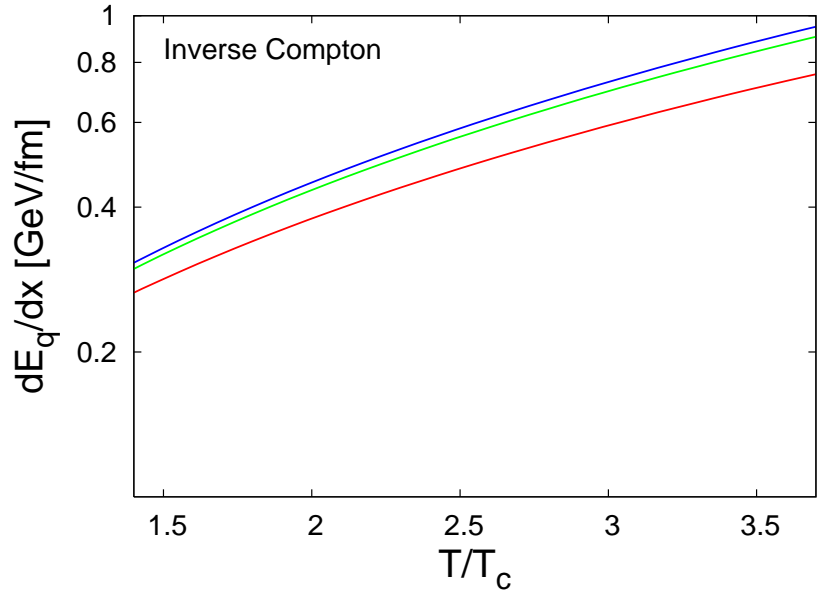}
\caption{\label{fig:dEdx_plot_IC} Energy loss of a light 
quark in QCD with $N_f =3$ massless quark flavors due to inverse Compton scattering 
as a function of $T/T_c$ for energies $E_q=20,\,100,\,200\;\text{GeV}$ of the 
incident quark (from bottom to top).  
The chiral phase transition temperature is given by $T_c\approx 161\;\text{MeV}$.}
\end{figure}

\begin{figure}
\includegraphics[clip,scale=0.9]{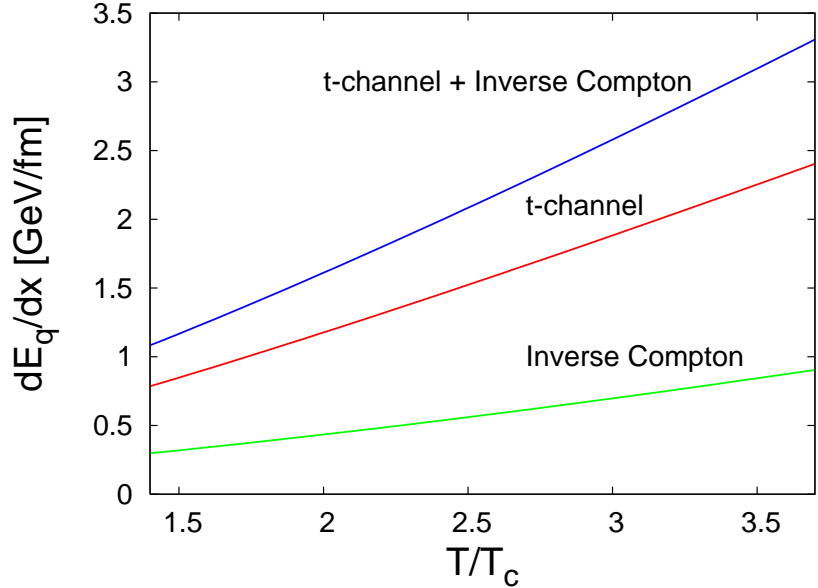}
\caption{\label{fig:dEdx_plot_COMPARE} Comparison of energy loss  
of a massless quark in QCD with $N_f =3$ massless quark flavors 
 due to inverse Compton scattering and due to $t$-channel gluon exchange  
as a function of $T/T_c$ for energies $E_q=100\;\text{GeV}$ of the 
incident quark. 
The chiral phase transition temperature is given by $T_c\approx 161\;\text{MeV}$.}
\end{figure}

Next, we demonstrate that quark-gluon inverse Compton scattering also  
gives a quantitatively relevant contribution to the energy loss for 
a quark traversing the quark-gluon plasma. 
Along the lines of the $t$-channel calculation we obtain for 
the energy loss due to IC-scattering for large $s$:
\be
\frac{dE^{\text{IC}}_q (T)}{dx}&=& \frac{8}{27}\pi\, T^2\,\int _{\mu _q ^2 (T)} ^{s}d|u|\,\left(\alpha _s ^2 (\sqrt{s},T)\frac{|u|}{s} +\alpha _s ^2 (\sqrt{|u|},T)\frac{s}{|u|}\right)\frac{s-|u|}{s^2}\,.
\label{eq:dEdxIC}
\ee
The results for energy loss due to this process as a function of $T/T_c$ for 
different energies of the incident quark $E_q$ are shown in Fig. \ref{fig:dEdx_plot_IC} 
for $N_f =3$ massless quark flavors.
Comparing the results for the $t$-channel process with the IC process in
Fig. \ref{fig:dEdx_plot_COMPARE}, we observe 
that the IC process enhances the energy loss of the incident quark.
Intuitively, this is clear since the s-u channel processes are central collisions where the high-energy particle can transfer more energy. 
However, due to the screening mass of the quark, the largest energy transfers are reduced.
As far as we know, this process has not yet been taken into account in previous 
theoretical work, cf.~\cite{Alam:2006qf}. 
Taking a temperature of $T\approx 2\,T_c$, then
the IC process gives $\frac{dE}{dx}\approx 0.43\;\text{GeV/fm}$ compared 
with $\frac{dE}{dx}\approx 1.15\;\text{GeV/fm}$ for the t-channel process.
The extra contribution due to the IC process helps to 
explain the energy loss observed at RHIC which is dominated
by the early stages of the collision where the plasma is dense and hot \cite{Zapp:2005kt}. 
The collisional energy loss is approximately of the same order as the radiative 
energy loss~\cite{Gyulassy:2003mc}, since the density of the scatterers is 
falling very quickly. For a recent discussion of this problem, see 
also Ref.~\cite{Alam:2006qf}.

\section{Conclusions and Outlook}\label{sec:conc}

We demonstrated that the running of the QCD coupling in the high-temperature plasma
is important. The coupling varies as a function of the temperature $T$ and the
resolution $k$. 
For momenta $|p|$ larger than the temperature $T$ the infrared growth of the 
coupling enhances the cross sections and the energy loss considerably compared 
with calculations with a constant value for the QCD coupling.
For momenta $|p|<T$, the coupling decreases due to the existence of a non-Gaussion 
infrared fixed point of the spatial $3d$ Yang-Mills theory. In fact, this 
decreasing coupling is only weakly sampled in the calculations because of the 
screening masses of the quarks and gluons which serve effectively as an IR cutoff.
In addition to the effects arising from the running of the QCD coupling, 
our results suggest that the inverse Compton process is a 
relevant mechanism for a high-energy quark to lose energy in the quark-gluon plasma. 
These findings may help to explain the observed energy loss observed at RHIC.

Based on our work presented here, one can estimate a thermalization time. One can 
approximate the kinetic equation as the 
scattering of  a test particle having a saturation momentum $Q_s$ 
($\approx 1.5\;\text{GeV}$ at RHIC) in a "plasma" of temperature $T_0$ 
($\approx 3\,T_c$) after the time $\tau=\tau _0$, 
when the colliding nuclei have passed through each other.
Explicitly considering that the expansion 
of the plasma leads to a decrease of $T$ as $T_0(\tau_0/\tau)^{1/3}$, one finds 
from collisional energy loss an equilibration time of $\tau\approx 4\tau_0$. 
This time is somewhat shorter than the time found from radiative energy 
loss~\cite{Moss:2001hq,Johnson:2000ph,Johnson:2001xf}.
Both mechanisms together seem to 
give a reasonable short thermalization time, $\tau _{\text{eff.}}\approx 3\,\tau_0$
without involving new mechanisms.
\acknowledgments
The authors are grateful to H. Gies, J. Stachel and K. Zapp for useful discussions. 
J. B. acknowledges financial support 
by the Gesellschaft f\"ur Schwer\-ionenforschung (GSI) Darmstadt.


\begin{thebibliography}{99}

\bibitem{Shuryak:2003xe}
E.~Shuryak,
Prog.\ Part.\ Nucl.\ Phys.\  {\bf 53}, 273 (2004)

\bibitem{Blaizot:2003tw}
  J.~P.~M.~Blaizot, E.~Iancu and A.~Rebhan,
  arXiv:hep-ph/0303185.

\bibitem{Braun:2005uj}
J.~Braun and H.~Gies,
arXiv:hep-ph/0512085; (submitted to Phys. Lett. B)
\bibitem{Braun:2006jd}
J.~Braun and H.~Gies,
JHEP {\bf 0606}, 024 (2006)
[arXiv:hep-ph/0602226].

\bibitem{Gies:2002af}
H.~Gies,
Phys.\ Rev.\ D {\bf 66}, 025006 (2002); 
{\bf 68}, 085015 (2003).

\bibitem{Abbott:1980hw}
L.~F.~Abbott,
Nucl.\ Phys.\ B {\bf 185}, 189 (1981).

\bibitem{Reuter:1993kw} 
M.~Reuter and C.~Wetterich,
Nucl.\ Phys.\ B {\bf 417}, 181 (1994); 
%
M.~Reuter and C.~Wetterich,
Phys.\ Rev.\ D {\bf 56}, 7893 (1997);
F.~Freire, D.~F.~Litim and J.~M.~Pawlowski,
Phys.\ Lett.\ B {\bf 495}, 256 (2000);

\bibitem{Bethke:2004uy}
  S.~Bethke,
  Nucl.\ Phys.\ Proc.\ Suppl.\  {\bf 135} (2004) 345.

\bibitem{DSE_Landau_gauge}
L.~von Smekal, R.~Alkofer and A.~Hauck,
Phys.\ Rev.\ Lett.\  {\bf 79}, 3591 (1997); 
%
D.~Atkinson and J.~C.~Bloch,
Mod.\ Phys.\ Lett.\ A {\bf 13}, 1055 (1998); 
%
C.~Lerche and L.~von Smekal,
Phys.\ Rev.\ D {\bf 65}, 125006 (2002); 
%
C.~S.~Fischer  and R.~Alkofer,
Phys. Lett. {\bf B536}, 177 (2002).
%
\bibitem{RG_Landau_gauge}
J.~M.~Pawlowski, D.~F.~Litim, S.~Nedelko and L.~von Smekal,
Phys.\ Rev.\ Lett.\  {\bf 93}, 152002 (2004);
%
C.~S.~Fischer and H.~Gies,
JHEP {\bf 0410}, 048 (2004).

\bibitem{Maas:2004se}
A.~Maas, J.~Wambach, B.~Gruter and R.~Alkofer,
Eur.\ Phys.\ J.\ C {\bf 37}, 335 (2004);
A.~Maas, J.~Wambach and R.~Alkofer,
Eur.\ Phys.\ J.\ C {\bf 42}, 93 (2005). 
\cite{Maas:2004se}

\bibitem{Peshier:2006ah}
A.~Peshier,
arXiv:hep-ph/0601119.

\bibitem{Pisarski:1987wc}
R.~D.~Pisarski,
Nucl.\ Phys.\ B {\bf 309}, 476 (1988).

\bibitem{Karsch:2000kv}
F.~Karsch, E.~Laermann and A.~Peikert,
Nucl.\ Phys.\ B {\bf 605} (2001) 579.

\bibitem{Gyulassy:2003mc}
M.~Gyulassy, I.~Vitev, X.~N.~Wang and B.~W.~Zhang,
arXiv:nucl-th/0302077.
%
\bibitem{Ellis:1991qj}
R.~K.~Ellis, W.~J.~Stirling and B.~R.~Webber,
Camb.\ Monogr.\ Part.\ Phys.\ Nucl.\ Phys.\ Cosmol.\  {\bf 8}, 1 (1996).

\bibitem {Bere}
V.S. Berezinskii, S.V. Bulanov, V. A. Dogiel, V.L. Ginzburg and V.S. Ptuskin;
Astrophysics of Cosmic Rays, p.124 ff.,Amsterdam , 1990

\bibitem {Aharonian} F. A. Aharonian; 
Very high energy cosmic gamma radiation : a crucial window on the extreme universe,
Singapore, 2004. 

\bibitem{Bjorken:1982tu}
J.~D.~Bjorken,
FERMILAB-PUB-82-059-THY

\bibitem{Peshier:2006hi}
A.~Peshier,
arXiv:hep-ph/0605294.

\bibitem{Alam:2006qf}
  J.~e.~Alam, A.~K.~Dutt-Mazumder and P.~Roy,
  arXiv:nucl-th/0608058.

\bibitem{Zapp:2005kt}
  K.~Zapp, G.~Ingelman, J.~Rathsman and J.~Stachel,
  Phys.\ Lett.\ B {\bf 637} (2006) 179
  [arXiv:hep-ph/0512300].
  
\bibitem{Moss:2001hq}
  J.~M.~Moss {\it et al.},
  arXiv:hep-ex/0109014.

\bibitem{Johnson:2000ph}
  M.~B.~Johnson {\it et al.}  [FNAL E772 Collaboration],
  Phys.\ Rev.\ Lett.\  {\bf 86}, 4483 (2001)
  [arXiv:hep-ex/0010051].

\bibitem{Johnson:2001xf}
  M.~B.~Johnson {\it et al.},
  Phys.\ Rev.\ C {\bf 65}, 025203 (2002)
  [arXiv:hep-ph/0105195].





\end{thebibliography}
\end{document}